\documentclass[pre,aps,twocolumn,showpacs]{revtex4-2}
\usepackage{epsf,graphics,graphicx,amsmath,color}
\usepackage[hidelinks]{hyperref}
\begin{document}


\title{Determining the purity of single-helical proteins from electronic specific heat measurements}

\author{Sourav Kundu}

\email{sourav.kunduphy@gmail.com}

\affiliation{Department of Physical Sciences, Indian Institute of Science Education and Research Kolkata, Mohanpur, West Bengal - 741246, India}

\author{Siddhartha Lal}

\email{slal@iiserkol.ac.in}

\affiliation{Department of Physical Sciences, Indian Institute of Science Education and Research Kolkata, Mohanpur, West Bengal - 741246, India}

\begin{abstract}

We present a theoretical investigation of the electronic specific heat (ESH) at constant volume ($C_v$) of single-helical proteins modeled within the tight-binding (TB) framework. We study the effects of helical symmetry, long-range hopping, environment and biological defects on thermal properties. We employ a general TB model to incorporate all parameters relevant to the helical structure of the protein. In order to provide additional insights into our results for the ESH, we also study the electronic density of states for various disorder strengths. We observe that the variation of the specific heat with disorder is very different in low and high temperature regimes, though the variation of ESH with temperature possesses a universal pattern upon varying disorder strengths related to  environmental effects. Lastly, we propose an interesting application of the ESH spectra of proteins. We show that by studying the ESH of single-helical proteins, one can distinguish a defective sample from a pure one. This observation can serve as the basis of a screening technique that can be applied prior to a whole genome testing, thereby saving valuable time $\&$ resources. 

\end{abstract} 


\maketitle

\section{Introduction}

Over the previous two decades, biomolecules have received considerable attention from physics and engineering communities because of their possible applications in nanoelectronics and spintronics devices~\cite{endres,zutic,genereux,cordes}. DNA has traditionally been a major focal point~\cite{kelley,fink,porath,cai,tran} in this regard, while other biomolecules (e.g., proteins) have not been studied in as much detail. This has, however, changed with recent progress in chirality induced spin selectivity (CISS) ~\cite{gohler,xie,guo_prl,guo_prb1,guo_pnas,galperin,eremko,yeganeh,guti,senthil,gersten,mishra,ben}, as both DNA and proteins have helical structures that can be used for efficient spin polarization. In 2011, Göhler {\it et al}.~\cite{gohler} showed that double-stranded DNA (ds-DNA) can be used as a good spin filtering agent with length dependent spin polarization up to 60$\%$. No spin polarization has, however, been achieved for single-stranded DNA (ss-DNA) till date. These findings have been theoretically supported by Guo {\it et al}.~\cite{guo_prl}. Recent experiments suggest that $\alpha$-helical proteins are also quite efficient in spin polarization process despite having a single helical structure~\cite{mishra,ben}, primarily due to the presence of multiple charge conduction pathways (MCCP). These results thus open up an opportunity to examine these single-helical structures from a new perspective, especially as different theoretical models have been proposed to explain these experimental results~\cite{guo_pnas}. Further, DNA is widely studied with respect its to electronic charge transfer properties, though different experimental results remain controversial~\cite{porath,cai,tran,zhang,storm,yoo,kasumov,conwell,dekker,ratner,beratan,sourav1,sourav3}. 
At the same time, charge transport through proteins has been studied in depth~\cite{jin,prytkova,beratan2,gao,sepunaru}. 
 
Indeed, much effort has been invested in understanding the transport properties of biomolecules, only a few studies exist of the thermal and thermodynamic properties of single-helical proteins~\cite{moreira1, sarmento, moreira2, moreira3}. Importantly, a recent study~\cite{mendes} showed that the knowledge of thermal properties of poly-peptides and similar biomolecules may be helpful in the determination of various neuro-degenerative diseases (e.g., Alzheimer $\&$ Parkinson). With this in mind, we conduct a detailed study of the electronic specific heat (ESH) of single-helical proteins. Our main aim is to investigate ESH spectra, and its variations with environmental effects and biological defects. Specifically, we apply tight-binding modeling to study the electronic specific heat of single-helical protein molecules. Employing a generic tight-binding framework to describe distinct structural properties of single-helical proteins, we study the effects of helical symmetry, long-range hopping, environment and biological defects on the ESH response of these helical biomolecules. We see that the ESH varies non-monotonically with temperature (as expected of a one-dimensional electron gas): it increases linearly at very low temperatures, rising sharply to a maxima within the low temperature range and then decaying exponentially as temperature is increased further. We have also observed that there exists an interplay between temperature and disorder in the ESH spectra, leading to a shift of the maxima ($(C_v)_{max}$) with increasing disorder.  

This paper is organized as follows. In Sec. II we introduce the model Hamiltonian and briefly describe our theoretical formulation. We analyze our numerical results in Sec. III, and conclude in Sec. IV.

\section{Model and Theoretical Formulation}

Following Ref.~\cite{guo_pnas}, we employ a generic tight-binding framework to model the single-helical proteins, with the tight-binding Hamiltonian for the protein molecules formed from the basis set spanned by the amino acids. Thus, the the tight-binding Hamiltonian for the protein molecule is given by
\begin{eqnarray}
& H_{pro}&= \sum\limits_{i=1}^N \epsilon_{i} c^\dagger_{i}c_{i}+ 
\sum\limits_{i=1}^{N-1}\sum\limits_{j=1}^{N-i} t_j c^\dagger_{i}c_{i+j}+\mbox{h.c.}~,
\end{eqnarray} 
where $c_i^\dagger$ ($c_i$) is the fermionic creation (annihilation) operator at the {\it i}th Wannier state of the protein molecule, $\epsilon_i=$ is the on-site potential energy of the amino acids at $i$th lattice site and $N$ is the total length of the molecule. Further, the $j$-th neighbouring hopping amplitude is given by $t_j$= $t_1 e^{-(l_j-l_1)/l_c}$, where $l_j$ is the distance between two neighbours $i$ and $i+j$, $l_c$ is the decay exponent and $t_1$ is the nearest neighbour hopping amplitude. In terms of the radius $R$, stacking distance $\Delta h$ and twisting angle $\Delta \phi$, we can write $l_{j}$ as~\cite{guo_prb2}
\begin{equation}
l_{j}=\sqrt{\left[2R \sin\left(\frac{j\Delta \phi}{2}\right)\right]^2 + \left(j\Delta h\right)^2}~.
\label{lnt}
\end{equation}
We have assumed here that, similar to the Slater-Koster scheme, all electronic wave functions decay exponentially over distance, and that the decay constant ($l_c$) can be obtained by matching with first-principle calculations~\cite{endres}. For a schematic representation, see Ref.~\cite{guo_pnas}.


The electronic specific heat at constant volume of proteins is determined by taking the first order derivative of average energy of the system with respect to temperature $T$
\begin{equation}
 C_v = \frac{\partial <E>}{\partial T}~,
\end{equation}
where 
\begin{eqnarray}
& <E>& = \sum\limits_{i=1}^N (E_i-\mu) f(E_i)~, \\
& f(E_i)&= \frac {1} {1+\exp(\frac{E_i-\mu}{k_BT})}~,
\end{eqnarray} 
and where $<E>$ is the average energy of the system, $E_i$ the energy of an electron at the {\it i}th eigenstate, $\mu$ is the chemical potential, $T$ is the temperature, $k_B$ is the Boltzmann constant and $f(E_i)$ is the occupation probability of the $i$-th eigenstate according to Fermi-Dirac statistics. Then, using the expressions of $<E>$ and $f(E_i)$, we find the following expression for electronic specific heat~\cite{sourav2} of single-helical proteins

\begin{equation}
 C_v = \sum\limits_{i=1}^N\frac{(E_i-\mu)^2 \exp(\frac{E_i-\mu}{k_B T})}{{k_B}T^2 (\exp(\frac{E_i-\mu}{k_B T})+1)^2}~.
\label{spht}
\end{equation}

Now, the ESH spectra is directly related to the electronic density of states (DOS) of a system. Therefore, to explain the ESH spectra, we also investigate the electronic density of states (DOS) of single-helical proteins. Employing the Green's function formalism~\cite{wu}, we find the average density of states (ADOS) of the system as  
\begin{equation}
 \rho(E) = - \frac{1}{N \pi} {\rm Im[Tr[G(E)]]}~,
\end{equation}
where 
$G(E)= (E-H+i\eta)^{-1}$ is the Green's function for the entire biomolecule with electron energy $E$, $\eta\rightarrow0^+$, $N$ the system length, $H$ the Hamiltonian of the biomolecule, and ${\rm Im}$ and ${\rm Tr}$ represent the imaginary part of the argument and the trace over the entire Hilbert space respectively. 

\section{Results and Discussions}

\begin{figure*}[htb]
\centering
\begin{tabular}{cc}
\includegraphics[width=0.42\textwidth]{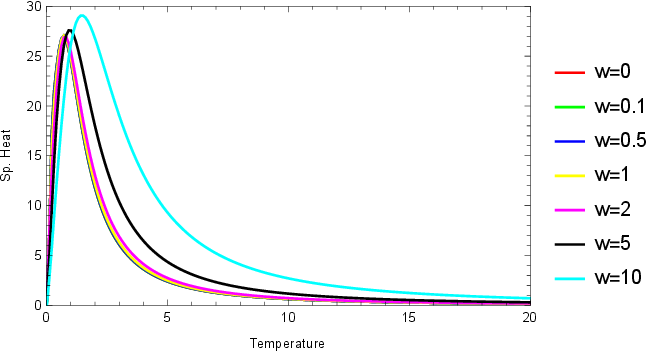}
\includegraphics[width=0.42\textwidth]{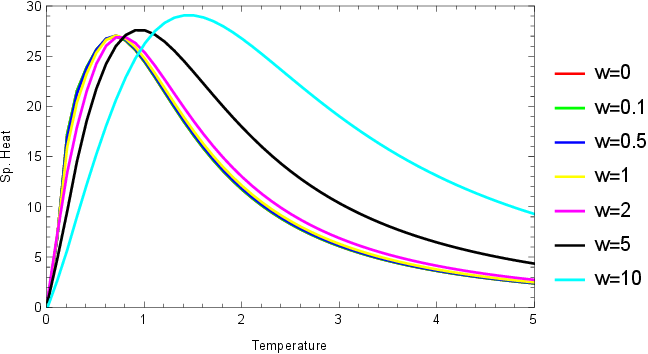}\\		
\end{tabular}
\caption{(Color online). Variation of electronic specific heat ($C_v$) of single helical proteins with 
temperature (T). Left panel shows the full scale with temperature varying upto 20 K. Right panel shows the zoomed in part of right panel within temperature range 0-5 K. Unit of w is in eV.}
\label{fig:1}
\end{figure*}

For numerical inspection of thermal properties in the absence of disorder, we set the on-site energies to a common value, $\epsilon_i$=0 eV~. We then incorporate the effects of various environmental fluctuations in terms of site-disorder, i.e., by considering an on-site energy $\epsilon_i$ that is randomly distributed within the range [$\epsilon_i$-w/2, $\epsilon_i$+w/2], where w represents the disorder strength. We use the following distance parameters in order to compute wave function overlap/ hopping amplitude ($t_j$) between two neighbouring amino acids (all units are in \AA{})~\cite{guo_pnas}: $l_1$=4.1, $l_2$=5.8, $l_3$=5.1, $l_4$=6.2, $l_5$=8.9, $l_6$=10.0 and $l_c$ is taken as 0.9. With these, we can calculate the related hopping integrals ($t_j$) which gives: $t_2$ $\sim$ 0.16$t_1$, $t_3$ $\sim$ 0.32$t_1$, $t_4$ $\sim$ 0.09$t_1$, $t_5$ $\sim$ 0.005$t_1$, $t_6$ $\sim$ 0.001$t_1$ and so on. It is clear that the $t_j$s decrease gradually, except for $t_3>t_2$, as $l_3<l_2$. We therefore restrict the range of hopping to $t_6$, and set $t_1$=$t$=1.0 eV. Finally, we set $k_B$=1.

\begin{figure*}[htb]
\centering
\begin{tabular}{ccc}
\includegraphics[width=0.32\textwidth]{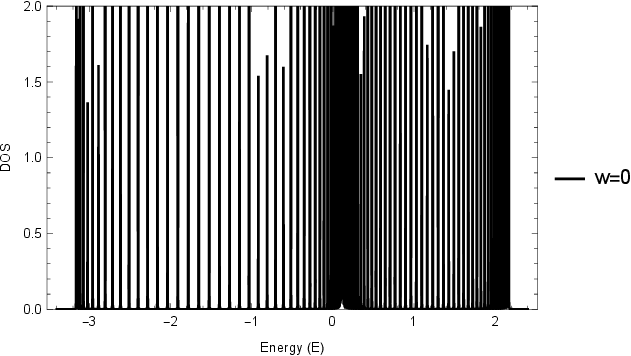}
\includegraphics[width=0.32\textwidth]{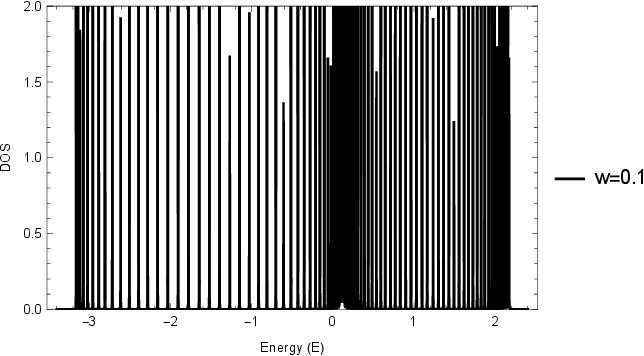}
\includegraphics[width=0.32\textwidth]{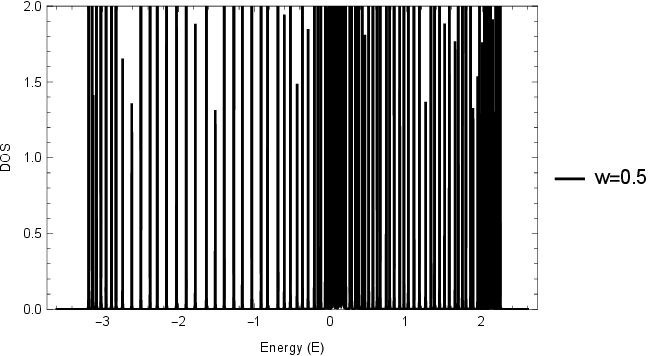}\\
\includegraphics[width=0.32\textwidth]{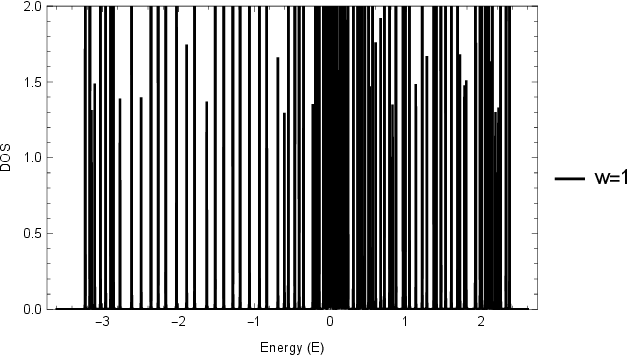}
\includegraphics[width=0.32\textwidth]{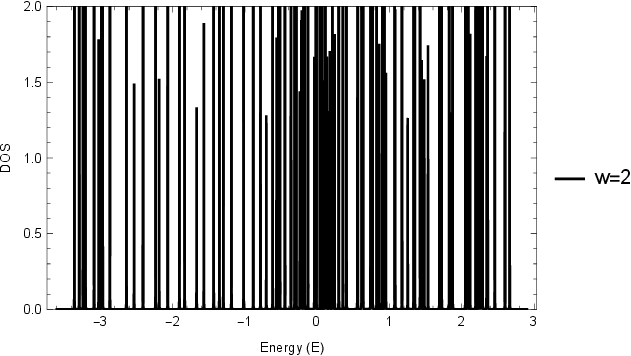}
\includegraphics[width=0.32\textwidth]{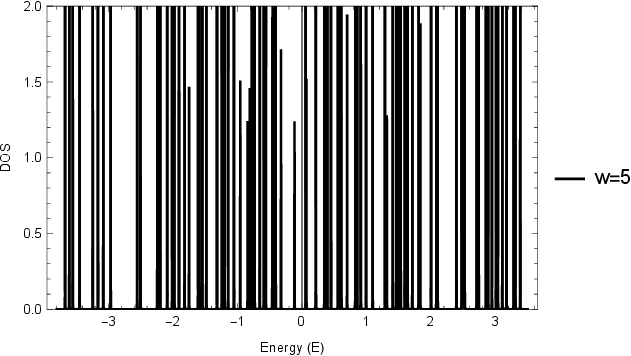}\\
\includegraphics[width=0.32\textwidth]{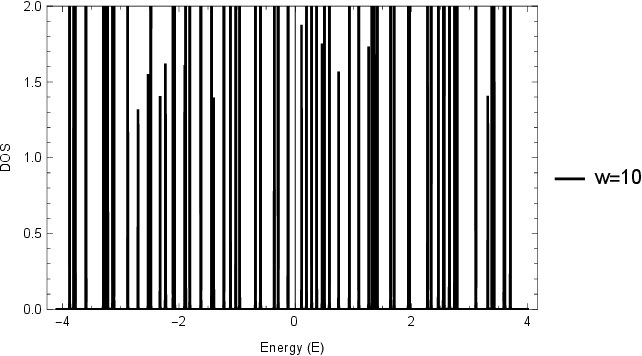}\\		
\end{tabular}
\caption{(Color online). Average Density of States (ADOS) in arbitrary units, vs. Energy (eV) for single helical proteins for various disorder strength (w). Unit of w is in eV.}
\label{fig:2}
\end{figure*}
In Fig.~\ref{fig:1}, we present the $C_v$ vs. temperature ($T$) at different disorder strengths (w). We first explain the nature of the $C_v$ vs temperature ($T$) curve. At low temperatures, only the states within the range $E_F\pm k_{B}T$ are accessible to the electrons in the tight-binding system, with $E_F$ being the Fermi energy. As the average energy of the system is given by $<E>$ = $\int E \rho(E) f(E) dE $, we can make the following approximations at low temperatures: dE~$\approx k_{B}T$, $\rho(E)\approx\rho$~= a constant, $f(E)~\approx$1. The average energy then becomes $<E>~\approx\rho k_{B}^2T^2$, leading to the specific heat $C_v$~= $\rho k_{B}^2T$ being proportional linearly to the temperature. 
On the other hand, $C_v$ falls exponentially with temperature at high temperatures (as can easily be seen from eq.\eqref{spht}). Further, the protein system is finite in length such that its energy spectra forms a band of finite width; at high temperatures, all states are easily accessible to the electrons. Consequently, the average energy $<E>$ becomes almost independent of temperature with increasing temperature, and ESH finally goes to zero in the very high temperature regime. The crossover between linear growth and exponential decay involves passage through a maximum. All of this is expected of a one dimensional gas of non-interacting electrons, and which can be modeled within the tight-binding framework.
We discuss below the effects of disorder. 

\begin{figure*}[htb]
\centering
\begin{tabular}{c}
\includegraphics[width=0.42\textwidth]{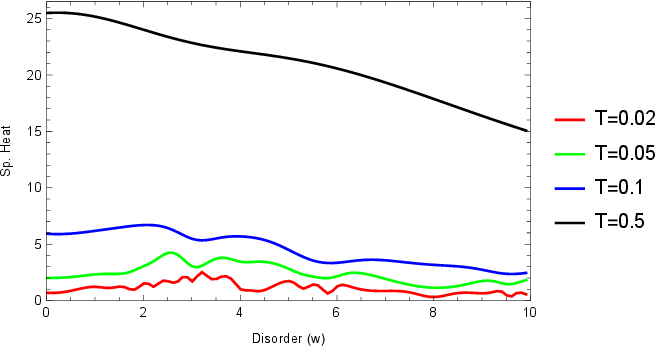}\\		
\end{tabular}
\caption{(Color online). Variation of electronic specific heat ($C_v$) of single helical proteins with 
environmental disorder (w) in low temperature (T) regime.}
\label{fig:3}
\end{figure*}

In Fig.~\ref{fig:3}, we plot the variation of specific heat $C_v$ with disorder in the low temperature range $T\le$ 0.5K. It is observed from the figure that $C_v$ decreases with disorder (w). The reason can be explained from the DOS profiles presented in Fig.~\ref{fig:2}. As we increase disorder from zero, new states appear as expected. Initially, these new states appear around the Fermi energy $(E_F)$. However, upon further increasing disorder, the band expands beyond the edges. This suggests that the new states are emergent around the band-edges; as the total number of states is fixed for the system, these new states at the band edges must arise at the cost of states around the $E_F$ (band-center). Thus, with increasing disorder (w), the DOS around $E_F$ must decrease. Given that only states around the band centre can be accessed at low temperatures, $C_v$ will thus decrease with increasing disorder (w). Some fluctuations in $C_v$ are also observed at low temperatures in Fig.\ref{fig:3}; this was previously reported~\cite{albu} for quasi-periodic DNA sequences, and we observe similar fluctuations in single-helical proteins due to environmental effects.

\begin{figure*}[htb]
\centering
\begin{tabular}{c}
\includegraphics[width=0.42\textwidth]{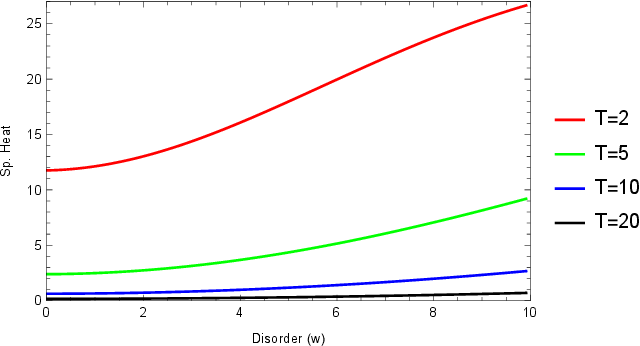}\\		
\end{tabular}
\caption{(Color online). Variation of electronic specific heat ($C_v$) of single helical proteins with 
environmental disorder (w) in high  temperature (T) regime.}
\label{fig:4}
\end{figure*}

In Fig.~\ref{fig:4}, we show the variation of $C_v$ with disorder strength (w) for the temperature range  T$\ge$2K), displaying clearly that $C_v$ increases with w. To explain these we once again look at the corresponding DOS profiles presented in Fig.~\ref{fig:2}. 
As discussed above, as we increase disorder, new states appear around the band-edges. Further, the energy of these new states also increases with increasing w. 
In turn, this leads an increase in the rate of change of average energy $<E>$ with T as we increase the disorder strength w. Consequently, $C_v$ increases with increasing w in the high temperature regime.

\begin{figure*}[htb]
\centering
\begin{tabular}{c}
\includegraphics[width=0.42\textwidth]{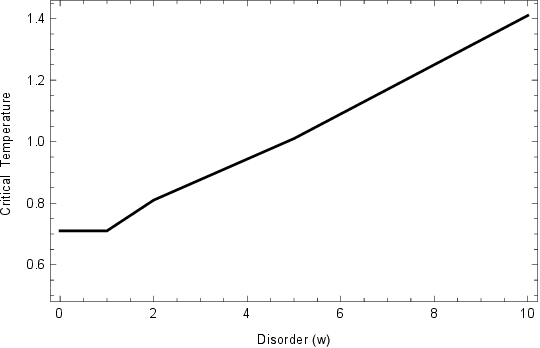}\\		
\end{tabular}
\caption{(Color online). Variation of critical temperature ($T_c$) of single helical proteins with 
environmental disorder (w). We can see that $T_c$ varies almost monotonically with w.}
\label{fig:5}
\end{figure*}

At this point, we make an important observation. In Fig.~\ref{fig:1}, 
we find that at low temperatures, $C_v$ decreases with increasing w at a given temperature. At high temperatures, on the other hand, $C_v$ is enhanced with w at a given temperature (as discussed above). Further, the peak of the $C_v$ vs. T curve ($(C_v)_{max}$, which determines the crossover temperature $T_c$) also increases and shifts towards higher temperatures as we increase disorder strength (w). In Fig.~\ref{fig:5}, we show the dependence of crossover temperature ($T_c$) with disorder strength (w).
The figure shows clearly a monotonic increases in $T_c$ with increasing w.  

\begin{figure*}[htb]
\centering
\begin{tabular}{c}
\includegraphics[width=0.42\textwidth]{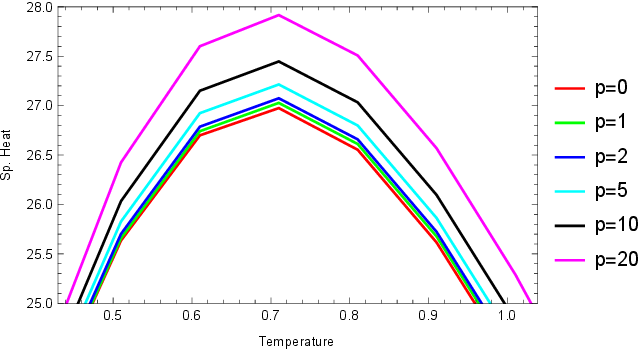}\\		
\end{tabular}
\caption{(Color online). Variation of ESH of single helical proteins in presence of biological defects. 
We can see as the strength (p denotes the percentage) of defects increases, the separations between characteristics maxima of ESH $(C_v)_{max})$ increases, making it more easier to distinguish a defective sample from a pure one.}
\label{fig:6}
\end{figure*}

In Fig.~\ref{fig:6}, we show that our observation of the increase of $(C_v)_{max}$ with increasing disorder (w) constitutes an important application for the determination of whether a proteins contains any defects or foreign elements. Traditionally, in order to sequence DNA or proteins, we rely on the Sanger~\cite{sanger} method based biomolecular sequencing; this is, however, costly and time consuming. An improvement could involve screening of a sample for defects prior to a whole genome sequencing. Here, we show that by measuring specific heat, one can obtain a first estimate of the purity of a given biological sample. 
We have considered several different concentrations of defects corresponding to the percentage of defective amino acid sites in a protein chain: $\sim$1$\%$, 2$\%$, 5$\%$, 10$\%$ and 20$\%$. 
The on-site energy of amino acids in a pure sample are set to $\epsilon_i=$0 eV, while for a defective site, it is set to 0.5 eV. In order to test the sensitivity of our proposed technique, we keep the defective site energy on the lower side (and closer to that of the pure sample). Clearly, if the method is able to detect small changes, we expect that it can detect more drastic changes as well. In Fig.~\ref{fig:6}, we plot $C_v$ vs temperature (T) curve and focus our attention around the peak ($(C_v)_{max}$) of the obtained curves. One can easily see that $(C_v)_{max}$ is found to increase significantly with even 1$\%$ concentration of defects. As the defect percentage increases, the increase in $(C_v)_{max}$ becomes even more vivid. In this way, by comparing the $(C_v)_{max}$ of a known pure sample with that obtained from a given test sample, one can easily confirm whether the test sample contains defects or not. In this way, one can preclude whole genome testing for a sample that does not contain any defects. In this way, one can eliminate the costs associated with unnecessary genome testing, a scenario made even welcome during an outbreak or pandemic. Further, the present results can easily be extended to the case of strong defects by setting the on-site energies to larger values. 

\section{Concluding Remarks}

The thermal properties of helical biomolecules are not well explored till date. There are very few results available on the specific heat of proteins~\cite{mendes,audun1,audun2}, and to the best of our knowledge, a complete study (including the roles played by helical symmetry and various other parameters) is missing.  Helical symmetry and longer-ranged nature of electron hopping are the very basic structural nuances of these biological units, because of which they are able to polarize electron spin. In this article, we have examined the electronic specific heat of single-helical proteins by using a tight-binding Hamiltonian incorporating all of its structural properties. Specifically, we have studied the effects of helical symmetry, long-range hopping, environmental effects and biological defects on the electronic specific heat of single-helical proteins. We observe a non-monotonic behaviour of $C_v$ with temperature; while this is consistent with that of a one-dimensional electron gas for the case of a uniform chain, this behaviour is observed to remain even in the presence of disorder. The variation of $C_v$ with disorder (w) is, however, quite different in low and high temperature range: increasing w decreases $C_v$ at low temperatures but increases it at higher temperatures. We have also shown that a knowledge of the ESH can help in detecting the presence of defects within a given protein. These defects can either be pathogens (i.e., foreign elements) or internal elements that can be attributed to the emergence of different neurological disorders (Alzheimer, Parkinsosn's) or even cancer. We find that a careful exploration of a global variable like ESH can provide significant insight into the purity of the composition of the protein sample: even a very small amount (~1$\%$) of defects can be detected in ESH spectra. Normally, alpha-helices can range between 4 to 40 amino acid residues, with the average length being 15 \AA{} (i.e., 10 residues) long. Here, we obtain visible separation in characteristic ESH for pure and defective samples for larger system sizes (e.g., $N=100$). In a realistic case (where the average protein length is typically one tenth of that considered by us), these effects are expected to be more prominent. This indicates that our proposed method can serve as a good basis for the formulation of new screening techniques of biological samples. Such a method can be especially effective in times of a pandemic, saving valuable time $\&$ resources. 

While theoretical explorations of thermal and thermodynamic properties of biomolecules are infrequent, experimental investigations are very rare~\cite{pranav}. To verify our theoretical predictions experimentally, heat exchange in bimolecular reactions e.g., protein binding, unfolding and ligand association etc. should be measured with considerable accuracy. There are only three techniques available for these type of measurements (see Ref.~\cite{jelesarov} for a detailed review): differential scanning calorimetry (DSC, which measures sample heat capacity with respect to a reference as a function of temperature), isothermal titration calorimetry (ITC, which measures the heat absorbed or rejected during a titration experiment) and thermodynamic calorimetry. Unfortunately, none of these techniques is presently able to separate the electronic contribution to the specific heat from other sources (e.g., vibrational modes etc.). However, we believe that our results are sufficiently interesting, and will spur efforts towards a detailed experimental investigation in the near future. 

\acknowledgments
SK thanks the CSIR, Govt. of India for funding through a Research Associate fellowship. SL thanks the SERB, Govt. of India for funding through MATRICS grant MTR/2021/000141 and Core Research Grant CRG/2021/000852.

\end{document}